\documentclass[aps,pre,groupedaddress]{revtex4}

\begin{document}

\title{Stable spinning optical solitons in three dimensions}
\author{D. Mihalache}
\altaffiliation{Institute of Solid State Theory and Theoretical Optics,
Friedrich-Schiller University Jena, Max-Wien-Platz 1, D-07743, Jena, Germany}
\altaffiliation{Department of Signal Theory and Communications,
Polytechnic University of Catalonia, ES 08034 Barcelona, Spain}
\author{D. Mazilu}
\altaffiliation{Institute of Solid State Theory and Theoretical Optics,
Friedrich-Schiller University Jena, Max-Wien-Platz 1, D-07743, Jena, Germany}
\author{L.-C. Crasovan}
\altaffiliation{Department of Signal Theory and Communications,
Polytechnic University of Catalonia, ES 08034 Barcelona, Spain}
\affiliation{Department of Theoretical Physics, Institute of Atomic Physics, 
PO Box MG-6, Bucharest, Romania}
\author{I. Towers}
\author{A. V. Buryak}
\affiliation{School of Mathematics and Statistics, University of New South Wales at ADFA,
Canberra, ACT 2600, Australia}
\author{B. A. Malomed}
\affiliation{Faculty of Engineering,
Tel Aviv University, Tel Aviv 69978, Israel}
\author{L. Torner}
\author{J. P. Torres}
\affiliation{Department of Signal Theory and
Communications,
Polytechnic University of Catalonia, ES 08034
Barcelona, Spain}
\author{F. Lederer}
\affiliation{Institute of Solid State Theory and Theoretical Optics,
Friedrich-Schiller
University Jena, Max-Wien-Platz 1, D-07743, Jena, Germany}

\begin{abstract}
We introduce spatiotemporal spinning solitons (vortex tori) of the
three-dimensional nonlinear Schr\"{o}dinger equation with focusing cubic and
defocusing quintic nonlinearities. The first ever found completely stable
spatiotemporal vortex solitons are demonstrated. A general conclusion is
that stable spinning solitons are possible as a result of competition
between focusing and defocusing nonlinearities.
\end{abstract}
\maketitle

Optical solitons (spatial, temporal, or spatiotemporal) are self-trapped
light beams or pulses that are supported by a balance between diffraction
and/or dispersion and various nonlinearities. They are ubiquitous objects in
optical media \cite{George}. Spatiotemporal solitons (STS)\ \cite{KanRub},
alias superspikes \cite{MBA,Manassah} or light bullets \cite{Yaron}, were
found in many works \cite{KanRub}~-~ \cite{Agrawal}. Although they cannot be
stable in the uniform Kerr ($\chi ^{(3)}$) medium \cite{collapse}, stability
can be achieved in saturable \cite{MBA,Edm}, quadratically nonlinear ($\chi
^{(2)}$) \cite{KanRub,quadr}, and graded-index Kerr media \cite{Agrawal}.
While a fully localized STS in three dimensions (3D) has not yet been found
in an experiment, 2D ones were observed in a bulk $\chi ^{(2)}$ medium \cite
{Wise}. The interplay of spatio-temporal coupling and nonlinearity may also
play an important role in self-defocusing media \cite{ManassahDefocusing}.

Spinning (vortex) solitons are also possible in optical media. Starting with
the works \cite{vortex}, both delocalized (``dark'') and localized
(``bright'') optical vortices in 2D were investigated \cite
{DiTrapp,unstable1,experiment}. In the 3D case they take the shape of a
torus (``doughnut'') \cite{Anton,new}. However, the only previously known
physical model which could support {\em stable} 3D vortex solitons is the
Skyrme model \cite{reviews}, which has recently found a new important
application to Bose-Einstein condensates (BEC) \cite{SkyrmeBEC}. Our
objective in this paper is to identify fundamental models of the
nonlinear-Schr\"{o}dinger (NLS) type in 3D that give rise to stable spinning
solitons, as NLS models are much simpler and closer to more experimental
situations, having applications to optics, BEC, plasmas, etc. (see below).

For bright vortex solitons stability is a major issue as, unlike their
zero-spin counterparts, the spinning solitons are prone to destabilization
by azimuthal perturbations. In 2D models with $\chi ^{(2)}$ and saturable
nonlinearities an azimuthal instability was revealed by simulations \cite
{unstable1} and observed experimentally \cite{experiment}. As a result, a
soliton with spin $1$ splits into two or three fragments, each being a
moving zero-spin soliton. Simulations of the 3D spinning STS in the $\chi
^{(2)}$ model also demonstrates its instability-induced splitting into
separating zero-spin solitons \cite{new}. Nevertheless, the $\chi ^{(2)}$
nonlinearity acting in combination with the self-{\em defocusing} Kerr ($%
\chi ^{(3)}$) nonlinearity, gives rise to the first examples of stable
spinning (ring-shaped) 2D solitons with spin $s=1$ and $2$ \cite{Isaac}. It
should be stressed that all the 2D spinning solitons actually represent
static spatial beams; on the contrary, 3D solitons are moving spatio{\em %
temporal} ones, which are localized not only in the transverse plane, but
also in the propagation coordinate, see below.

A model which may support stable spinning solitons in 3D is the one with a
cubic-quintic (CQ) nonlinearity, which (in terms of optics) assumes a
nonlinear correction to the medium's refractive index in the form $\delta
n=n_{2}I-n_{4}I^{2}$, $I$ being the light intensity. The CQ nonlinearity was
derived, starting from the Maxwell-Bloch equations, for light propagation
combining resonant interaction with two-level atoms and dipole interactions
between the atoms \cite{dipole}, or the Kerr nonlinearity of a waveguide 
\cite{John} (see also Ref. \cite{Manassah-twolevel}). A unifying feature of
those media is competition between different nonlinearities, and stable
solitons may exist in the range of intensities where the competition takes
place (which may be controlled, for instance, through the density of
two-level atoms in a $\chi ^{(3)}$ waveguide). In fact, the NLS equation of
the CQ type is a generic model, which also applies to Langmuir waves in
plasmas\cite{plasma} and BEC \cite{BEC} (although in the latter case,
three-body interactions, which give rise to the quintic term, may induce
losses through recombination of the colliding atoms, thus making the quintic
coefficient complex).

In the first simulations of 2D solitons with spin $1$ in the CQ model, it
was found that they propagated in a stable way, provided that their energy is not too small 
\cite{Q}. A later analysis, based on the computation of linear-stability
eigenvalues, demonstrated that some of the spinning 2D solitons considered
in Ref. \cite{Q} are subject to a weak azimuthal instability. Nonetheless,
in another part of their existence region, with very large energies,
solitons with spin $s=1$ and $s=2$ were confirmed to be truly stable in the
2D CQ model \cite{IsaacPLA} (all the solitons with $s\geq 3$ are unstable).

It was recently shown by direct simulations of the CQ model \cite{new2} that
3D spinning solitons with moderate energies are unstable against azimuthal
perturbations, while the ones with very large energies, i.e., broad
``doughnuts'' with a small hole in the center, were robust under
propagation. However, a consistent stability analysis makes it necessary to
compute eigenvalues of small perturbations. We will conclude that
sufficiently broad STS with spin $s=1$ are stable, the stability region
occupying $\approx 20\%$ of their existence region, while all the STS with $%
s\geq 2$ are unstable.

The evolution of the electromagnetic field envelope $A$ in the dispersive CQ
medium is governed by the NLS equation, 
\begin{equation}
2i\kappa _{0}A_{z}+\nabla _{\bot }^{2}A+\kappa _{0}DA_{\tau \tau }+2\kappa
_{0}^{2}(n_{2}/n_{0})|A|^{2}A-2\kappa _{0}^{2}(n_{4}/n_{0})|A|^{4}A=0,\,\tau
\equiv t-z/V,  \label{Egn}
\end{equation}
where $z$ and $t$ are the propagation coordinate and time, $\kappa _{0}$ and 
$V$ are the propagation constant and group velocity of the carrier wave, $%
D>0 $ is the temporal dispersion, $\nabla _{\bot }^{2}$ acting on the
transverse coordinates $x$ and $y$. Equation (\ref{Egn}) does not include
higher-order effects, such as self-steepening, stimulated Raman scattering,
non-paraxial diffraction, and third-order temporal dispersion (which, in
another context, were taken into regard for spatiotemporal superspikes in
Refs. \cite{Manassah}), as we anticipate that only broad solitons (with the
temporal width $\sim 1$ ps), for which these effects are small, may be
stable.

Defining rescaled variables $u=\sqrt{n_{4}/n_{2}}A$, $T=n_{2}\sqrt{2\kappa
_{0}/Dn_{0}n_{4}}\tau $, $Z=\left( \kappa _{0}n_{2}^{2}/n_{0}n_{4}\right) z$%
, and $(X,Y)=\kappa _{0}n_{2}\sqrt{2/n_{0}n_{4}}\left( x,y\right) $, we
transform Eq. (\ref{Egn}) into a normalized form \cite{new2}, 
\begin{equation}
iu_{Z}+\left( u_{XX}+u_{YY}+u_{TT}\right) +|u|^{2}\,u\,-|u|^{4}\,u\,=0.
\label{scaled}
\end{equation}
STS solutions to Eq. (\ref{scaled}) are sought as $u=U(r,T)\exp (is\theta
)\exp (i\kappa Z)$, where $r$ and $\theta $ are the polar coordinates in the
transverse plane, $\kappa $ is a propagation constant parameterizing the
family of solutions sought for, and $s$ is an integer spin. The real
amplitude $U$ obeys the equation 
\begin{equation}
\left( U_{rr}+r^{-1}U_{r}-s^{2}r^{-2}U+U_{TT}\right) -\kappa
U+U^{3}\,-U^{5}\,=0,  \label{stat}
\end{equation}
supplemented by the condition that $U$ must decay exponentially as $%
r\rightarrow \infty $ and $T\rightarrow \infty $ (due to the definition of $%
\tau $ in Eq. (\ref{Egn}), the latter condition implies that the solution's
snapshot taken at $t={\rm const}$ is localized in the propagation coordinate 
$z$ - in fact, exactly the same way as 1D solitons are localized in optical
fibers \cite{Agr}).

Equation (\ref{scaled}) conserves the energy $E~=~\int \int \int \left|
u(X,Y,T)\right| ^{2}dXdYdT$, Hamiltonian 
\begin{equation}
H=\int \int \int \left[
|u_{X}|^{2}+|u_{Y}|^{2}+|u_{T}|^{2}-(1/2)|u|^{4}+(1/3)|u|^{6}\right] dXdYdT,
\label{inv2}
\end{equation}
momentum (equal to zero for the solutions considered), and angular momentum
in the transverse plane, $L=\int \int \int \left( \partial \phi /\partial
\theta \right) |u|^{2}dXdYdT$, where $\phi $ is the phase of the complex
field $u$. Relations between $L$, $H$ and $E$ for a stationary spinning
soliton follow from Eq. (\ref{stat}): $L=sE$; $H=\kappa E-\frac{2}{3}\int
\int 2\pi rU^{6}(r,T)drdT$ \cite{new2}.

We have numerically found families of 3D spinning solitons with a toroidal
shape. To quantify the solutions, in Fig. 1 we show the propagation constant 
$\kappa $ and $H$ vs. $E$ for both $s=0$ and $s=1,2$ solitons. They exist
for $E$ exceeding a threshold value, which increases with $s$. The full and
dashed lines in Fig. 1 correspond to stable and unstable branches according
to results presented below. The $s=0$ branch of the solutions is divided
into stable and unstable portions on the basis of the known criterion which
states that the fundamental ($s=0$) soliton branch undergoes a change in the
stability where $dE/d\kappa =0$ \cite{VK}. However, this criterion ignores
azimuthal instability, which is frequently fatal for spinning solitons.

The most revealing information on the stability of solitons is provided by
analysis of a linearized version of Eq. (\ref{scaled}). To this end, we seek
perturbation eigenmodes of the general form, 
\begin{eqnarray}
u(Z,r,T,\theta )-U(r,T)\exp \left[ i(s\theta +\kappa Z)\right] &=&f(r,T)\exp
\left \{ \lambda _{n}Z+i[(s+n)\theta +\kappa Z]\right \}  \nonumber \\
&&+g^{\ast }(r,T)\exp \left \{ \lambda _{n}^{\ast }Z+i[(s-n)\theta +\kappa
Z]\right \} \,,  \label{perturbation}
\end{eqnarray}
where $n>0$ is an arbitrary integer azimuthal index of the perturbation, $%
\lambda _{n}$ is the (complex) instability growth rate sought for, and the
functions $f$ and $g$ obey equations 
\begin{eqnarray}
i\lambda _{n}f+\frac{\partial ^{2}f}{\partial T^{2}}+\frac{\partial ^{2}f}{%
\partial r^{2}}+r^{-1}\frac{\partial f}{\partial r}-(s+n)^{2}r^{-2}f-\kappa
f+\left( 2-3U^{2}\right) U^{2}\,f+\left( 1-2U^{2}\right) U^{2}g &=&0, 
\nonumber \\
-i\lambda _{n}g+\frac{\partial ^{2}g}{\partial T^{2}}+\frac{\partial ^{2}g}{%
\partial r^{2}}+r^{-1}\frac{\partial g}{\partial r}-(s-n)^{2}r^{-2}g-\kappa
g+\left( 2-3U^{2}\right) U^{2}\,g+\left( 1-2U^{2}\right) U^{2}f &=&0.
\label{growthrate}
\end{eqnarray}
The solutions must decay exponentially at $r\rightarrow \infty $, and vanish
as $r^{\left| s\pm n\right| }$ at $r\rightarrow 0$.

To solve Eqs. (\ref{growthrate}), we used a known numerical procedure \cite
{unstable1,Akh}, which produces results presented in Fig. 2. The most
persistent unstable eigenmode has $n=2$, for both $s=1$ and $s=2$. As is
seen in Fig. 2, with the increase of $\kappa $, the instability of the
soliton with $s=1$, accounted for by ${\rm Re\,}\lambda _{2}$, disappears at 
$\kappa =\kappa _{{\rm st}}\approx 0.13$, and the stability region extends
up to $\kappa =\kappa _{{\rm offset}}^{{\rm (3D)}}\approx 0.17$,
corresponding to infinitely broad solitons (which implies that the vortex of
the dark-soliton type \cite{vortex}, that may be regarded as an infinitely
broad spinning soliton, is stable too). The relative width of the stability
region is $\left( \kappa _{{\rm offset}}^{{\rm (3D)}}-\kappa _{{\rm st}%
}\right) /\kappa _{{\rm offset}}^{{\rm (3D)}}\approx 0.2$. However, there is 
{\em no} stability region for 3D solitons with $s=2$, in contrast to the 2D
vortex solitons in the CQ model \cite{IsaacPLA}. In the case when a spinning
soliton is unstable, its instability is {\it oscillatory}; the corresponding
frequency, ${\rm Im}\lambda $, is of the same order of magnitude as ${\rm Re}%
\lambda $ at the maximum-instability point (see Fig. 2), and $\lambda $
becomes purely imaginary at $\kappa =\kappa _{{\rm st}}$.

The above results were checked in direct simulations of Eq. (\ref{scaled})
by means of the Crank-Nicholson scheme combined with the Gauss-Seidel
iteration procedure. In Fig. 3 we show the amplitude and energy vs. $Z$ for
the soliton with $s=1$, generated by two different initial configurations,
with the same energy of the initial configuration, $E_{0}=$ $13070$.
Robustness of the spinning STS is attested to by the fact that it can be
generated from a Gaussian with a nested vortex whose shape is far from the
soliton's exact form. (Energy loss evident in Fig. 3(b) is caused by
emission of radiation in the course of the formation of the stable STS;
naturally, the loss is larger for the initial Gaussian configuration, which
is farther from the exact soliton's shape.) Figure 4 shows the gray-scale
contour plots of the intensity and phase of both the input Gaussian with a
nested vortex and emerging spinning STS at $Z=400$.

The instability of the $s=2$ solitons is illustrated in Figs. 5 and 6. The
azimuthal instability breaks them into zero-spin solitons which fly out
tangentially, relative to the circular crest of the original soliton. It is
noteworthy that, at an early stage of the evolution shown in Fig. 5, the
spinning soliton splits into two fragments, in accordance with the
growth-rate calculations predicting the dominant instability to be against
the $n=2$ perturbation mode, but the subsequent nonlinear evolution is more
involved. In Fig. 6(b), the spinning STS splits into three fragments with 
{\em unequal} energies. The fragmentation of this soliton (having $\kappa
=0.09$) into three parts is in accordance with the growth-rate calculation,
see Fig. 2(b).

In conclusion, we have found the first example of stable three-dimensional
spinning solitons in a dispersive medium which combines cubic and quintic
nonlinearities. Only sufficiently broad solitons with spin $s=1$ may be
stable. However, the existence of stable spinning 3D solitons is a generic
fact, as it is not limited to the cubic-quintic nonlinearity: our
preliminary studies indicate that these stable physical objects may also
occur in media with competing quadratic and self-defocusing cubic
nonlinearities. In fact, a condition which turns out to be necessary
for the existence of stable spinning solitons is a competition between two
different nonlinearities, one focusing and the other one defocusing.

\begin{acknowledgments}
D. Mihalache, D. Mazilu and L.-C. Crasovan acknowledge support from the
Deutsche Forschungsgemeinschaft (DFG) and European Community (Access to
Research Infrastructure Action of the Improving Human Potential Program).
B.A.M. and I.T. appreciates support from the Binational (US-Israel) Science
Foundation through the grant No. 1999459. I.T. and A.V.B. acknowledge
financial support of the Australian Research Council. L.T. and J.P.T.
acknowledge support by TIC2000-1010.
\end{acknowledgments}

\bibliography{basename of .bib file}

\begin{figure}
\caption{The propagation constant $\protect\kappa$ (a) and Hamiltonian $H$
(b) of the 3D spinning soliton vs. its energy $E$.}
\label{Fig.1}
\end{figure}

\begin{figure}
\caption{The growth rate of perturbations, ${\rm Re}\protect\lambda$, with
different values of the azimuthal number $n$ (indicated by labels near the
curves) vs. the soliton's propagation constant $\protect\kappa$: (a) $s=1$;
(b) $s=2$.}
\label{Fig. 2}
\end{figure}

\begin{figure}
\caption{Evolution of the maximum amplitude  of the soliton with $s=1$ (a) and 
its energy (b), as generated by initial configurations in the form of a 
Gaussian with a nested vortex (continuous curves) or a torus close to the 
stationary spinning soliton corresponding to $\protect\kappa =0.15$ 
(dashed curves).}
\label{Fig.3}
\end{figure}

\begin{figure}
\caption{The formation of the soliton with spin $s=1$: (a) the initial
Gaussian with a nested vortex; (b) its phase field; (c) the spinning soliton
at $Z=400$; (d) the phase field at $Z=400$. A cross section of the fields at 
$T=0$ is shown.}
\label{Fig.4}
\end{figure}

\begin{figure}
\caption{Gray-scale plots showing the developing instability of the spinning
soliton with $s=2$ and $\protect\kappa=0.13$ at $Z=0$ (a), $Z=600$ (b), 
$Z=620$ (c), and $Z=640$ (d).}
\label{Fig.5}
\end{figure}

\begin{figure}
\caption{Isosurface plots illustrating the instability of the $s=2$ soliton
with $\protect\kappa=0.09$: (a) $Z=0$, (b) $Z=250$.}
\label{Fig.6}
\end{figure}
\end{document}